\documentstyle[aps,epsf]{revtex} 

\begin{document} 
  
\centerline{\large\bf Center vortex model for the infrared sector of}
\vspace{0.2cm}
\centerline{\large\bf Yang-Mills theory -- Confinement and 
Deconfinement}

\bigskip
\centerline{M.~Engelhardt\footnote{Supported by DFG 
under contract En 415/1--1.} 
and H.~Reinhardt\footnote{Supported in part by DFG 
under contract Re 856/4--1.} }
\vspace{0.2 true cm} 
\centerline{\em Institut f\"ur Theoretische Physik, Universit\"at 
T\"ubingen }
\centerline{\em D--72076 T\"ubingen, Germany}
  
\begin{abstract}
A model for the infrared sector of Yang-Mills theory based on magnetic
vortices represented by (closed) random surfaces is investigated 
using lattice Monte Carlo methods. The random surfaces are governed 
by a surface area action and a curvature action. The model generates
a finite-temperature deconfinement transition; the coupling constants 
of the model can be chosen such as to reproduce the SU(2) Yang-Mills
ratio of the deconfinement temperature to the square root of the
zero-temperature string tension, $T_c /\sqrt{\sigma_{0} } =0.69$.
This yields a physical trajectory in the space of coupling constants
on which the confinement properties are approximately invariant. An
at first sight surprisingly accurate prediction of the spatial string
tension in the deconfined phase results, which can be made plausible in 
view of the specific space-time structure of the vortex configurations in 
this phase. The confinement properties are shown to be intimately tied 
to the percolation properties of the vortex surfaces.
\end{abstract}

\vskip .5truecm
PACS: 12.38.Aw, 12.38.Mh, 12.40.-y

Keywords: Center vortices, infrared effective theory, confinement,
deconfinement transition
\medskip

\section{Introduction}
The infrared sector of strong interaction physics is characterized
by nonperturbative phenomena. Most prominent among these are the
confinement of quarks and gluons into color-singlet hadrons, and
the spontaneous breaking of chiral symmetry, which, through the
associated (quasi-) Goldstone bosons, i.e. the pions, dominates the
low-lying hadronic spectrum. On the theoretical side, there is
presently convincing evidence from numerical lattice calculations
\cite{rothe} that Yang-Mills theory, i.e. QCD without
dynamical quark degrees of freedom, indeed generates confinement.
Concomitantly, diverse physical pictures of the QCD vacuum have been 
proposed which generate a confining potential between color sources, 
among others, the dual Meissner effect mechanism \cite{tho81},\cite{mag}, 
random magnetic vortices \cite{thoo}-\cite{spag}, the stochastic vacuum 
\cite{dosch}, the leading-log model of Adler \cite{adler}, and dual QCD 
\cite{zacha}. Other models, for instance instanton models 
\cite{inst}, have foregone a description of confinement 
and concentrated on generating spontaneous chiral symmetry breaking. 
Furthermore, there is compelling evidence from lattice Monte Carlo 
experiments \cite{rothe} that, as temperature is raised, one 
encounters a transition to a deconfined phase of Yang-Mills theory in 
which colored constituents can propagate over distances much larger than
typical hadronic sizes.

As the above listing already indicates, there presently exists a
disparate collection of model explanations for different
nonperturbative QCD phenomena, but not a consistent, comprehensive
picture of the degrees of freedom dominating the infrared sector.
Perspectives for bridging this gap have recently arisen in the
framework of the {\em magnetic vortex picture} of the QCD vacuum.
In this picture, initially explored in \cite{thoo}-\cite{spag},
the Yang-Mills functional integral is assumed to be dominated by 
disordered vortex configurations. These vortices represent closed magnetic 
flux lines in three-dimensional space; correspondingly, their world 
sheets in four-dimensional space-time are two-dimensional. They should 
be contrasted with the electric flux degrees of freedom encoded by 
Wilson lines. In fact, they can be regarded as dual to the latter. 
Wilson loops are sensitive to the (quantized) magnetic flux carried 
by the vortices; conversely, one can think of closed vortices as 
measuring the electric flux carried by a Wilson line.

On a space-time lattice, this relation is particularly manifest
in the fact that magnetic fluxes are defined on the lattice which is dual
to the one on which Wilson lines are defined (two lattices being dual
to one another means that they have the same lattice spacing $a$,
but are displaced from one another by the vector $(a/2,a/2,a/2,a/2)$).
In continuum language, using magnetic degrees of freedom means
switching from the usual canonical variables, namely vector potential
and electric field, to magnetic field variables along with appropriate
canonical conjugates. In the magnetic language, the constraint analogous
to Gau\ss ' law is the Bianchi identity, which enforces continuity of
magnetic flux. In order to emphasize the dual relation between vortices 
and electric flux, the above terminology, originating from the canonical
framework, will be used in the following even when discussing the 
corresponding objects covariantly in 3+1 dimensions, i.e. including 
their (Euclidean) time evolution. The duality between magnetic and
electric fluxes in particular provides a heuristic motivation for using
vortices to describe the infrared regime of Yang-Mills theory. Whereas
electric degrees of freedom become strongly coupled in the infrared,
leading to the nonperturbative effects highlighted above, it conversely
seems plausible to expect magnetic degrees of freedom to be weakly
coupled\footnote{For Yang-Mills theory, no explicit duality 
transformation which manifestly exchanges strong and weak coupling 
regimes is known. However, in related theories, e.g. supersymmetric
extensions \cite{seiberg}, such a transformation can be constructed.}.
They can thus be hoped to furnish an adequate representation
for the true infrared excitations of the theory.

Early evidence that magnetic vortices may indeed form in the Yang-Mills
vacuum came from the observation that a constant chromomagnetic field
is unstable with respect to the formation of tubular domains. This led
to the formulation of the Copenhagen ``spaghetti'' vacuum
\cite{spag}. Due to its technical complexity, this 
approach largely concentrated on local properties of vortices, as
opposed to their global topological character, which will take on
a decisive role in the present work.
Also in the lattice formulation of Yang-Mills theory, different
possibilities of defining vortices were explored 
\cite{mack},\cite{tomold}. On this track, a number of encouraging
developments have recently taken place; in particular, a practicable
procedure for isolating and localizing vortices in lattice gauge
configurations has been constructed \cite{deb96}-\cite{giedt}
using an appropriate gauge\footnote{For a recent discussion on the
meaning of this gauge-fixing procedure and possible alternatives and
generalizations, see \cite{tombmay}-\cite{vfind}. This discussion 
further underscores the need to independently explore the physics
of vortices, e.g. via the model presented in this work, beyond the
technical discussion of how they can be identified in lattice gauge
configurations.}. This has made it possible to gather a 
wealth of information about these collective degrees of freedom, 
using lattice experiments, which was previously inaccessible. Vortices 
appear to dominate long-range gluonic physics not only at zero 
temperature \cite{deb97}, but also at finite temperatures 
\cite{temp},\cite{tlang} and are able to generate both the confined as 
well as the deconfined phases. Moreover, there are indications that also 
chiral symmetry breaking is induced by vortices 
\cite{forc1},\cite{forc2},\cite{forc3} and that vortices are potentially
suited to account for the topological susceptibility of the Yang-Mills
ensemble \cite{cont},\cite{cornwall},\cite{forc1},\cite{preptop}.

If, however, the vortex picture of infrared Yang-Mills dynamics is to 
attain practical value beyond a qualitative interpretation of the 
lattice results, it must be ultimately developed into a full-fledged 
calculational tool, with a simplified model dynamics of the vortices 
allowing to concentrate on the relevant infrared physics. The work 
presented here is intended as an initial step in this direction. On the one 
hand, the effective vortex model proposed below is shown to qualitatively 
reproduce the confinement aspects of Yang-Mills theory, including the
finite-temperature transition to a deconfined phase; on the other hand, 
it is verified that the parameters of the model can be chosen such as to 
quantitatively replicate the relation between deconfinement temperature
and zero-temperature string tension known from lattice Yang-Mills
experiments. Thematically, the present report is restricted to the
confinement characteristics of the theory; because of this, only one
genuine prediction will be presented, namely of the behavior of the
spatial string tension in the deconfined phase. Another test of the 
predictive power of the model is discussed in a companion paper
\cite{preptop}, which focuses on the topological susceptibility of the
vortex surface ensemble proposed below. The model, which is
formulated below for the case of SU(2) color, is open to many 
refinements in its details; however, in its present form it is 
entirely adequate to reproduce the vortex phenomenology hitherto 
extracted from lattice Yang-Mills simulations. In particular,
the finite-temperature deconfinement transition can be understood 
in terms of a transition between two phases in which the vortices either 
percolate throughout (certain slices of) space-time or not.

\section{Vortex model}
\label{modsect}
The SU(2) vortex model under investigation in this work is defined by the
following properties:
\vspace{0.2cm}

{\bf Vortices multiply any Wilson loop by a factor corresponding to
a nontrivial center element of the gauge group whenever they pierce 
its minimal area.} Center elements of a group are those elements which 
commute with all elements in the group; e.g. in the case of SU(2), one 
has the trivial phase $1$ and one nontrivial element, namely the phase 
$-1$. An SU(2) color vortex thus contributes 
a factor $-1$ to a Wilson loop when it pierces the minimal area of the 
latter; its flux is quantized. This can be viewed as the defining 
property of a vortex \cite{thoo}-\cite{mack} beyond any specific model
assumption about its dynamics such as presented below. It specifies
how vortices couple to electric fluxes, in particular the fluxes
implied by the world lines of quark sources. This property provided
the original motivation for the proposal of the magnetic vortex
picture of confinement \cite{thoo}-\cite{corn79}. If the vortices
are distributed in space-time sufficiently randomly, then samples of
the Wilson loop of value $+1$ (originating from loop areas pierced an
even number of times by vortices) will strongly cancel against samples
of the Wilson loop of value $-1$ (originating from loop areas pierced
an odd number of times by vortices), generating an area law fall-off.
The circumstances under which this heuristic argument is valid will be
one of the foci of the present work. As a last remark, the following
should be noted. Above, the value a Wilson loop takes in a given
vortex configuration was defined via the number of times the minimal
area spanned by the loop is pierced by vortices. One should not
misinterpret this as an arbitrariness. Due to the closed
nature of the vortices, any other choice of area results in the same
value for the Wilson loop. If one wishes to formulate the influence of 
vortices on Wilson loops in a manner which does not refer to any spanning 
of the loop by an area, then it is the linking number between the vortices 
and the loops which determines the value of the latter. For practical 
purposes, however, the minimal area specification is very convenient. 
For the purpose of finite temperature studies, note that the above 
specification applies equally to the area spanned by two Polyakov loops.
\vspace{0.2cm}

{\bf Vortices are closed two-dimensional random surfaces in 
four-dimensional (Euclidean) space-time.} Vortex surfaces will be modeled as 
consisting of plaquettes on a four-dimensional space-time lattice. In a 
given lattice configuration, a plaquette can be associated with two values, 
0 or 1. The value 0 indicates that the plaquette in question is not part 
of a vortex, whereas the value 1 indicates that it is part of a vortex.
Note that this lattice is dual to the one on which ``electric'' 
degrees of freedom would be defined, such as Yang-Mills link
variables, or the associated Wilson loops. How vortices emerge on the 
dual lattice in the framework of Yang-Mills theory via center gauge fixing 
and center projection is discussed in detail in \cite{deb96}-\cite{giedt}. 
The fact that vortices are closed will be implemented in the following
way in numerical Monte Carlo experiments. Only Monte Carlo updates are
allowed which simultaneously change the values of
all the plaquettes making up the surface of an (arbitrary) 
three-dimensional elementary cube in the four-dimensional lattice. This 
can be interpreted as the creation of a vortex in the shape of the 
cube surface on the lattice. Such an algorithm generates only 
closed two-dimensional surface configurations\footnote{Note that this
construction also reflects that vortex surfaces can be viewed as boundaries 
of three-dimensional volumes in four-dimensional space-time \cite{cont}.}. 
Note that if a given plaquette which is being updated was already part of a 
vortex before the update, then it ceases to be part of a vortex after the 
update. One can think of this in terms of two vortices annihilating each 
other on a plaquette, if they both occupy that plaquette\footnote{For SU(2) 
color, there is only a single non-trivial center element $Z=-1$. As a 
consequence, two vortices annihilate, since $Z^2 =1$. For higher SU(N) 
groups, where the center elements are defined by $Z^N =1$, superposition 
of two vortices will in general yield a residual vortex flux, introducing 
the possibility of vortex branchings.}. The net result is that the 
plaquette is not part of any vortex.
\vspace{0.2cm}

{\bf Vortex surfaces are associated with a physical transverse thickness.}
In order to represent regular, finite action, configurations in 
Yang-Mills theory, vortices must possess a physical transverse
thickness in the directions perpendicular to the vortex surface.
This thickness has furthermore been argued to be of crucial
phenomenological importance, e.g. in generating the approximate
Casimir scaling behavior of Wilson loops in the adjoint representation
of the gauge group at intermediate distances \cite{fa97},\cite{fa98}.
A finite vortex thickness in particular implies that it does not make
sense to consider configurations in which two parallel vortex surfaces
are closer to each other than the vortex thickness; i.e., vortices cannot 
be packed arbitrarily densely. This feature is implemented in the present
vortex model simply via the lattice spacing, which forces parallel
vortices to occur a certain distance from each other. This should not be 
misconstrued to mean that the vortices have a hard core. Rather, when two 
parallel vortices come too close to each other, their fluxes can be 
considered to annihilate in the sense that their superposition is considered
equivalent to the vacuum. The reader is reminded that this is precisely
the manner in which the effect of a Monte Carlo update on a vortex
configuration was specified above. The algorithm thus reflects very
closely the underlying physical picture. Note furthermore that the
implementation of the vortex thickness via the lattice spacing implies
that the latter is treated as a fixed physical scale. This will be
commented upon in more detail further below. As a last remark, it
should be noted that at this stage, no explicit transverse profile
for the vortices has been introduced, as would be necessary e.g. for
the purpose of correctly describing the behavior of adjoint
representation Wilson loops \cite{fa97},\cite{fa98}. The vortex
thickness enters only via the minimal distance between parallel vortex 
surfaces, whereas the surfaces themselves are still treated as
infinitely thin. The introduction of an explicit transverse profile
is one of the possible refinements of the model presented here.
\vspace{0.2cm}

{\bf Vortices are associated with an action density per surface area.}
This is reflected in a corresponding explicit term in the action, 
cf. eq. (\ref{epsdef}) below.
\vspace{0.2cm}

{\bf Vortices are stiff.}
While an ultraviolet cutoff on the space-time fluctuations of the
vortex surfaces is already implied by the lattice spacing, vortices
will be endowed with a certain stiffness beyond this via an explicit term 
in the action, cf. eq. (\ref{cdef}) below. Specifically, if two plaquettes
which share a link but do not lie in the same two-dimensional plane
are both part of a vortex, then this fact will be penalized with
a certain action increment. Thus, vortices are stiffer than implied
by the lattice spacing alone, and the stiffness can be regulated with
an independent parameter.
\vspace{0.2cm}

\section{Physical Interpretation of the Model}
\label{sectphys}
Before investigating the properties of the model vortices defined
above, it is necessary to clarify more precisely which degrees of
freedom these vortices are to represent. The infrared structure
of typical Yang-Mills configurations is thought to be encoded in
thick magnetic vortices. The term ``thick'' means that, whereas
vortices on large scales form two-dimensional surfaces in 
four-dimensional space-time, they possess an extension, i.e. a
regular profile function, in the directions perpendicular to the
surface. This extension is one of the quantities determined in
the framework of the Copenhagen vacuum \cite{spag} and also has been 
probed by lattice experiments \cite{deb97},\cite{giedt},\cite{corr}.
As already mentioned in the previous section, it is instrumental in
explaining the approximate Casimir scaling behavior of Wilson loops 
in the adjoint representation of the gauge group at intermediate 
distances \cite{fa97},\cite{fa98}.

By contrast, the center projection vortices which arise in the framework 
of the maximal center gauges \cite{deb96}-\cite{giedt} are (in the 
continuum limit \cite{cont}) infinitely thin surfaces which provide a 
rough localization of the thick vortices described above, as has
also been ascertained empirically using lattice experiments
\cite{deb97},\cite{giedt}. It is possible to define the center
projection vortices on arbitrarily fine lattices. Thus, while their
effective action after integrating out all other Yang-Mills degrees
of freedom presumably does already contain the QCD scale $\Lambda_{QCD} $, 
it does not yet describe an infrared effective theory; center projection
vortices may still exhibit fluctuations of arbitrarily short wavelengths,
within the profile of the thick vortices they represent. The QCD scale
in the center projection vortex effective action merely controls on 
which scale this action becomes nonlocal. As one diminishes the lattice 
spacing, the nonlocality of the effective action presumably becomes 
more and more pronounced\footnote{Note that the center projection
vortex effective theory, which on the lattice is a $Z(2)$ gauge
theory with non-standard action, must contrive to avoid the
deconfining transition well-known to occur in the standard $Z(2)$
gauge theory with plaquette action as the coupling is diminished
in approaching the continuum limit.}.

\begin{figure}[t]
\centerline{
\epsfxsize=6cm
\epsffile{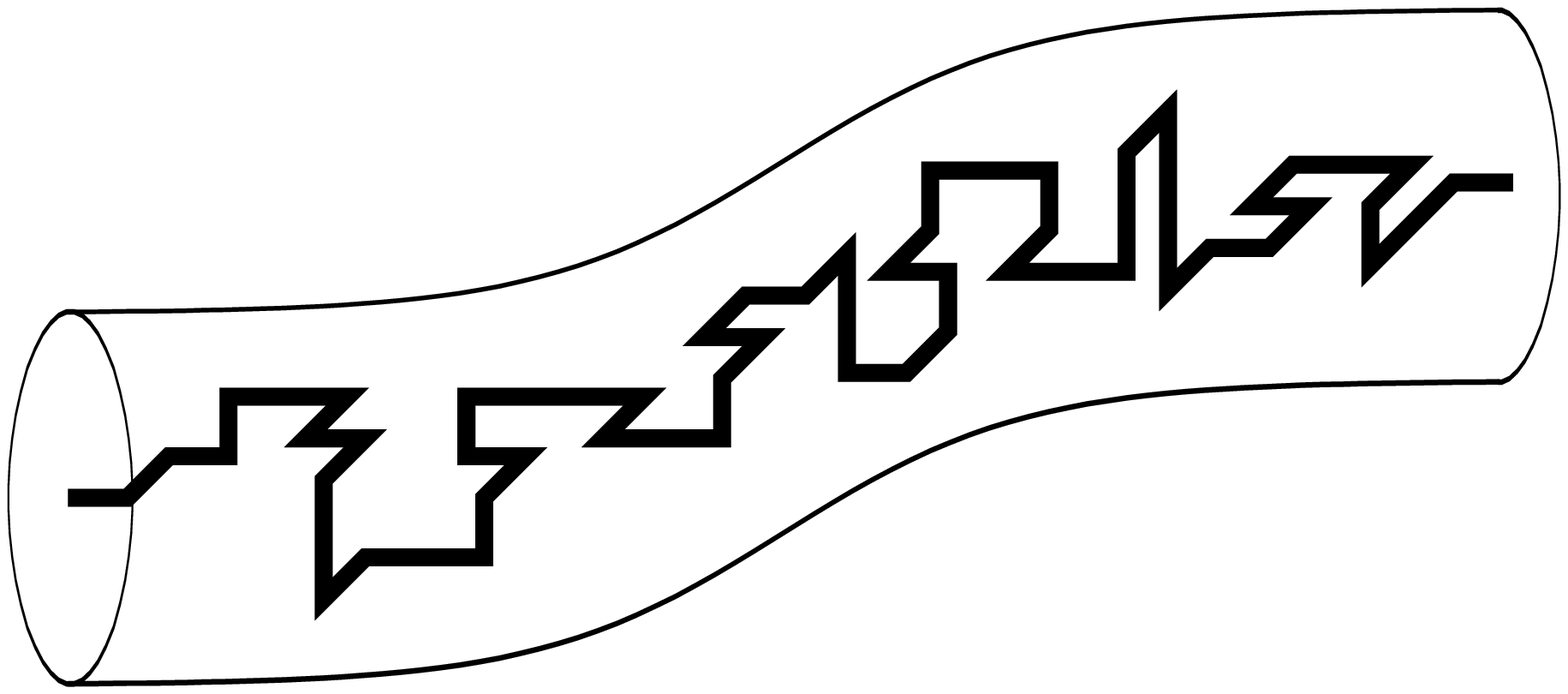}
\hspace{2cm}
\epsfxsize=6cm
\epsffile{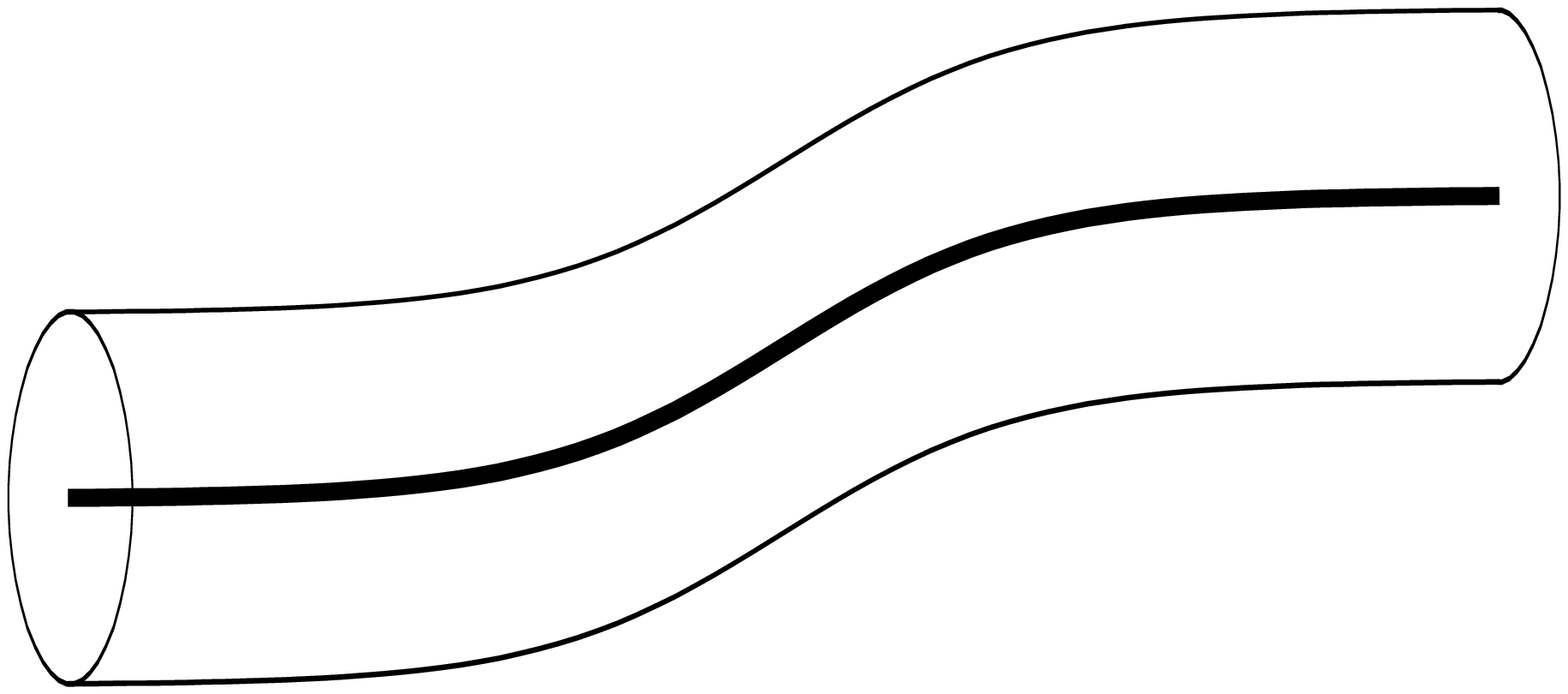}
}
\vspace{-1cm}
\caption{Thick vortices (in a three-dimensional slice of space-time)
associated with rough lattice center projection vortices (left), and
smooth thin model vortices (right), respectively.}
\label{flucpv}
\end{figure}

Evidence for the aforementioned short wavelength fluctuations has been 
gathered e.g. in \cite{corr}, where the binary correlations between 
center projection vortex intersection points with a given space-time 
plane were measured. This correlation function, while exhibiting the 
renormalization group scaling expected of a physical quantity, appears 
to diverge at small distances. Such behavior can be understood from 
short wavelength fluctuations of the center projection vortices as 
follows: Consider a plane which cuts a thick vortex along a (smeared-out) 
line, and consider furthermore intersection points of the associated center 
projection vortex with this plane, cf. Fig. \ref{flucpv}. Due to the 
transverse fluctuations of the projection vortices, one will find a 
strongly enhanced probability of detecting such intersection points 
close to one another (compared with the probability one would expect 
from the mean vortex density).
\vspace{-3cm}

The precise location of a center projection vortex within the profile
of the associated thick vortex is gauge-dependent; it changes
as different specific realizations of the maximal center gauge
are adopted \cite{cont},\cite{forc3}. The vortex model proposed in this 
work does not aim to reproduce these gauge-dependent fluctuations. It
is intended as a true low-energy effective theory; the model vortex
surfaces defined in the previous section, while formally thin, are
meant to represent the center of the profile of a thick vortex, 
which is smooth on short length scales, without reproducing the 
short-wavelength fluctuations of the corresponding center projection 
vortex, cf. Figure \ref{flucpv}. Alternatively, the model vortices can 
be interpreted as low-energy effective degrees of freedom obtained by 
integrating out, or smoothing\footnote{A specific smoothing procedure
applied to center projection vortices in Yang-Mills theory was investigated
in \cite{bertle}.} over, all ultraviolet fluctuations in the 
abovementioned center projection vortex effective theory, down to some 
fixed physical scale (encoded in the vortex model via the lattice spacing).
Note that such a procedure also eliminates the complicated nonlocal 
dynamics of the center projection vortices and leaves a vortex model 
action which should be well described by a few local terms, in the 
spirit of a gradient expansion \cite{cont}. It is this conceptual framework 
which led to the vortex model postulated in the previous section.
Note that this picture in particular implies that one should not
expect the center projection vortex density measured in lattice Yang-Mills
experiments, which does obey the proper renormalization group scaling law
\cite{langf} (note erratum in \cite{temp}), \cite{giedt}, to match the 
density of the model vortices. Rather, the latter should be substantially 
lower.

In view of this framework, it becomes clear that the lattice spacing
enters the vortex model as a fixed physical cutoff scale. In other words,
it is not envisaged to take the continuum limit of the model by
reducing the lattice spacing and accordingly renormalizing the
coupling constants. If one wishes to formulate the model in the
continuum, then the lattice spacing must be replaced by some other
fixed physical cutoff scale related to the thickness of the vortices.
Thus, the lattice model is only insofar a caricature of continuum
vortex physics as it merely allows the vortices to run parallel to the 
space-time axes. One could in principle remedy this e.g. 
by representing the vortex surfaces as triangulations in space-time, 
presumably again with some lower bound on the areas of the elementary 
triangles.

Accordingly, the lattice spacing introduces specific physical effects
into the vortex model, as already mentioned in the previous section.
On the one hand, the spacing accounts for aspects of the finite vortex 
thickness by preventing parallel vortex surfaces from occurring too 
close to one another. On the other hand, it introduces an ultraviolet 
cutoff on the space-time fluctuations of the vortices, which however 
is superseded by an explicit curvature penalty in the action. It must 
be emphasized that these effects enter the model not as unwanted lattice
artefacts, but as a specific realization of physical features also 
present in any corresponding continuum picture of vortices.

\section{Formal definition}
The properties of the vortex model described above can be summarized 
formally as follows. The basic variables are plaquettes
$p_n^{\{ \mu , \nu \} } $ on a four-dimensional space-time lattice,
extending from a lattice site described by the four-vector $n$
into the (positive) $\mu $ and $\nu $ directions. Note that the
superscripts $\{ \mu ,\nu \} $, where always $\mu \neq \nu $, are
unordered sets, i.e. there is no distinction between $\{ \mu ,\nu \} $
and $\{ \nu ,\mu \} $. The variables $p_n^{\{ \mu , \nu \} } $ can take 
the values 0 or 1. Furthermore, in the following, $e_{\mu } $
will denote the vector in $\mu $ direction of the length of one lattice
spacing. The partition function reads
\begin{equation}
Z = \left( \prod_{n} \prod_{\mu ,\nu \atop \mu < \nu } \ \ 
\sum_{p_n^{\{ \mu , \nu \} } =0}^{1} \right)
\Delta [\, p_n^{\{ \mu , \nu \} } ] \exp (-S [\, p_n^{\{ \mu , \nu \} } ] )
\label{partit}
\end{equation}
with the constraint
\begin{eqnarray}
\Delta [\, p_n^{\{ \mu , \nu \} } ] &=&
\prod_{n} \prod_{\mu } \delta_{L_{n}^{\mu } \bmod 2 ,0} \\
L_{n}^{\mu } &=& \sum_{\nu \atop \nu \neq \mu } \left(
p_n^{\{ \mu , \nu \} } + p_{n-e_{\nu } }^{\{ \mu , \nu \} } \right)
\end{eqnarray}
enforcing closedness of the vortex surfaces by constraining the number
$L_{n}^{\mu } $ of occupied plaquettes attached to the link extending from 
the lattice site $n$ in $\mu $ direction to be even, for any $n$ and $\mu $. 
How this constraint is conveniently enforced in Monte Carlo simulations was
described in section \ref{modsect}. The action consists of a surface area
part and a curvature part, $S = S_{area} + S_{curv} $,
which read
\begin{eqnarray}
S_{area} &=& \epsilon \sum_{n} \sum_{\mu , \nu \atop \mu < \nu }
p_n^{\{ \mu , \nu \} } \label{epsdef} \\
S_{curv} &=& \frac{c}{2} \sum_{n} \sum_{\mu } \sum_{\nu , \lambda \atop 
\nu \neq \mu , \lambda \neq \mu , \lambda \neq \nu } 
\left( p_n^{\{ \mu , \nu \} } p_n^{\{ \mu , \lambda \} }
+ p_n^{\{ \mu , \nu \} } p_{n-e_{\lambda } }^{\{ \mu , \lambda \} }
+ p_{n-e_{\nu } }^{\{ \mu , \nu \} } p_n^{\{ \mu , \lambda \} }
+ p_{n-e_{\nu } }^{\{ \mu , \nu \} }
p_{n-e_{\lambda } }^{\{ \mu , \lambda \} } \right) \label{cdef} \\
&=& \frac{c}{2} \ \sum_{n} \sum_{\mu } \left[
\left( \sum_{\nu \atop \nu \neq \mu } \left(
p_n^{\{ \mu , \nu \} } + p_{n-e_{\nu } }^{\{ \mu , \nu \} } \right) 
\right)^2 -\sum_{\nu \atop \nu \neq \mu }
\left( p_n^{\{ \mu , \nu \} } + p_{n-e_{\nu } }^{\{ \mu , \nu \} } 
\right)^2 \right] \nonumber
\end{eqnarray}
While the lower expression for $S_{curv} $ is more compact, the upper
expression exhibits its construction more clearly: For every link
extending from the lattice site $n$ in $\mu $ direction, all pairs
of attached plaquettes whose two members do not lie in the same plane
are considered and, if both members are part of a vortex, this is
penalized with an action increment $c$.

It should be noted that this type of random surface action has in 
recent times also attracted interest in connection with a rather
different physical motivation than the one espoused in the present 
work \cite{savvi},\cite{bilke}. These investigations correspondingly 
focus on entirely different observables associated with the random 
surfaces. In particular, only what would be interpreted as the 
zero-temperature case in the vortex framework is considered, whereas
the treatment below emphasizes also the generalization to finite
temperatures and the resulting phenomena.

Wilson loops are determined by the number of times
vortices pierce their minimal area. As a generic example, consider
a rectangular Wilson loop with corners defined as follows. Let
$n_0 $ be an arbitrary lattice site and $m_0 =n_0 +
(e_1 +e_2 +e_3 +e_4 )/2$. Place the corners of the Wilson loop at
$ \{ m_0 , m_0 +Je_1 , m_0 + Ke_2 , m_0 + Je_1 + Ke_2 \}$ with
integer $J,K$; as already mentioned in section \ref{modsect}, Wilson
loops are defined on the lattice dual to the one the vortices are
constructed on. Then, in any given vortex configuration, the number
of times the minimal area spanned by this Wilson loop is pierced by 
vortices is
\begin{equation}
Q = \sum_{j=1}^{J} \sum_{k=1}^{K} p_{n_0 +je_1 +ke_2 }^{ \{ 3,4\} }
\end{equation}
and the Wilson loop consequently takes the value $W=(-1)^{Q} $.

\section{Survey of the plane of coupling constants}
\label{survsect}
\begin{figure}
\centerline{
\epsfxsize=10cm
\epsffile{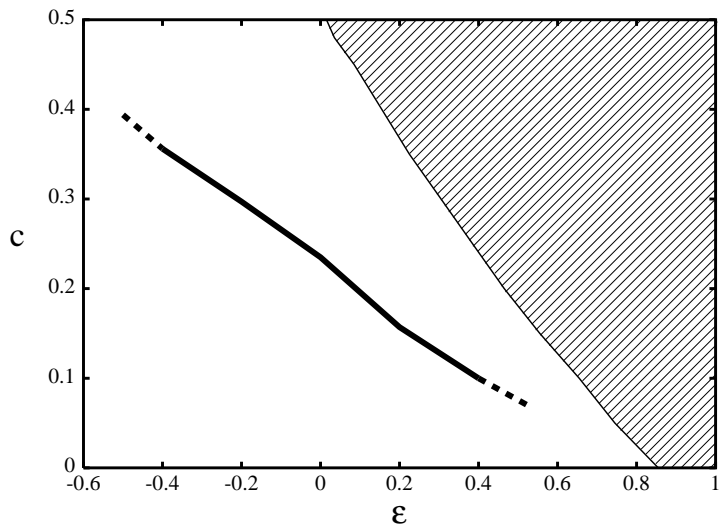}
}
\caption{Plane of coupling constants $\epsilon ,c$. The non-confining
region (at zero temperature) has been shaded. The thick solid line in
the confining region locates pairs of $\epsilon ,c$ at which the
Yang-Mills ratio $T_c /\sqrt{\sigma_{0} } \approx 0.69$ is reproduced
by the vortex model.}
\label{fig2}
\end{figure}

Carrying out measurements of Wilson loops of different sizes on a
symmetric ($16^4 $) lattice, one finds that the plane of coupling
constants $\epsilon ,c$ can be partitioned into a confining and a
non-confining region, cf. Fig. \ref{fig2}. Since the symmetric lattice 
constitutes an approximation to the zero-temperature (i.e. infinite 
lattice) theory (the physical size of the lattice spacing will be 
determined further below), the confining sector is of course the one of 
interest.

Furthermore, for a large range of coupling constants $\epsilon ,c$ in
the confining region, one finds a deconfinement phase transition when
raising the temperature of the system by decreasing the number of
lattice spacings $N_t $ making up the Euclidean time direction and
evaluating the heavy quark potential using Polyakov loop correlators.
Since the lattice only provides a discrete set of temperatures for given 
lattice spacing, it is necessary to use an interpolation procedure to 
determine the deconfinement temperature. For fixed $\epsilon $, the 
curvature coefficient $c$ was varied and the values of $c$ were recorded 
at which the inverse deconfinement temperature crosses the values
$a,2a$ and $3a$ ($a$ denoting the lattice spacing). Interpolation
then allows to define the ratio $T_c /\sqrt{\sigma_{0} } $ for all $c$,
where $T_c $ is the deconfinement temperature and $\sigma_{0} $ is the
zero temperature string tension. For the purpose of modeling SU(2)
Yang-Mills theory, it is desirable to reproduce the value
$T_c /\sqrt{\sigma_{0} } \approx 0.69$, cf. \cite{teper}. The line in the 
coupling constant plane for which this is achieved is displayed in 
Fig. \ref{fig2}. The dotted end of the line for $\epsilon > 0.4$ indicates 
that the $T_c /\sqrt{\sigma_{0} } \approx 0.69$ trajectory was not explored
further into this direction, because string tension measurements became
too noisy to allow its accurate determination; however, the authors
have no evidence that the trajectory stops at any particular point before
(presumably) ultimately reaching the non-confining region. On the other
hand, in the region $\epsilon < -0.4$, the system appears to become 
unstable; for $\epsilon = -0.6$, unphysical oscillatory behavior of the 
potential between static sources can be clearly observed. This is not
surprising, since for $\epsilon < 0$, the model action ceases to be
manifestly positive. Only for a certain limited region of negative
$\epsilon $ can one expect the model to be stabilized by the cutoff on 
the vortex density implied by the lattice spacing.

Setting the scale by positing a zero-temperature string tension of 
$\sigma_{0} = (440\mbox{MeV} )^2 $, measurements of $\sigma_{0} a^2 $ 
allow to extract the lattice spacing $a$. The results obtained on the
$T_c /\sqrt{\sigma_{0} } \approx 0.69$ trajectory for different
values of $\epsilon $ are summarized in Table \ref{table0}. 
The lattice spacing only varies by about 10 \% on the aforementioned
trajectory. This corroborates the interpretation of the 
lattice spacing as a fixed physical quantity discussed at length
in section \ref{sectphys}. It must again be emphasized that the role
of the lattice spacing in the vortex model is fundamentally different
e.g. from its role in lattice gauge theory. There, the lattice spacing
represents an unphysical regulator, and physical quantities must be
extrapolated to the continuum limit, where a certain scaling behavior
connecting the coupling constant and the lattice spacing arises due to
the scale invariance of the classical theory. On the other hand, in the
vortex model, the $T_c /\sqrt{\sigma_{0} } \approx 0.69$ trajectory
also implies a type of scaling behavior, but one which connects the
coupling constants $\epsilon $ and $c$. The lattice spacing $a$ by
contrast must be counted among the physical quantities, which are to
be accorded a fixed value, just like the string tension or the
deconfinement temperature. Indeed, as discussed in section \ref{sectphys},
the lattice spacing has a definite interpretation connected with the
transverse thickness of the vortices. It is reassuring that the
lattice spacing in fact does behave accordingly on the
$T_c /\sqrt{\sigma_{0} } \approx 0.69$ trajectory by remaining
approximately constant.

Most of the measurements in the subsequent sections refer specifically
to the case $\epsilon =0$; by constraining the model to the
$T_c /\sqrt{\sigma_{0} } \approx 0.69$ trajectory, this implies using
the set of coupling constants
\begin{equation}
\epsilon = 0 \ \ \ \ \ \ \ c = 0.24 \ .
\label{choice}
\end{equation}
When other choices of coupling constants are used, this is explicitly
stated. In particular, in the next section, a prediction of the
spatial string tension in the deconfined phase will be presented;
there, measurements at different points along the 
$T_c /\sqrt{\sigma_{0} } \approx 0.69$ trajectory will be taken to
demonstrate the stability of the prediction.

As a final remark, note that the lattice spacing $a$ also specifies
the ultraviolet limit of validity of the effective vortex theory.
In the case of the choice of coupling constants (\ref{choice}), one has
$a=0.39$ fm, cf. Table \ref{table0}, which corresponds to a maximal 
momentum representable on the lattice of 
$\Lambda = \pi /a \approx 1600 \mbox{MeV} $.
\vspace{0.7cm}

\begin{table}[h]

\[
\begin{array}{|c||c|c|c|}
\hline
\epsilon & \sigma_{0} a^2 & a & 1/a \\
\hline\hline
0.4 & 0.87 & 0.42 \, \mbox{fm} & 1.55 \, T_c \\ \hline
0.2 & 0.84 & 0.41 \, \mbox{fm} & 1.58 \, T_c \\ \hline
0 & 0.755 & 0.39 \, \mbox{fm} & 1.67 \, T_c \\ \hline
-0.2 & 0.73 & 0.38 \, \mbox{fm} & 1.70 \, T_c \\ \hline
-0.4 & 0.70 & 0.375 \, \mbox{fm} & 1.73 \, T_c \\ \hline
\end{array}
\]
\vspace{0.3cm}

\caption{Lattice spacing $a$ for different $\epsilon $ on the
$T_c /\sqrt{\sigma_{0} } \approx 0.69$ trajectory, cf. Fig. \ref{fig2}.
The scale is set by positing a zero-temperature string tension of
$\sigma_{0} = (440\mbox{MeV} )^2 $.}

\label{table0}

\end{table}

\section{Confinement and percolation}

\begin{figure}
\centerline{
\epsfxsize=12cm
\epsffile{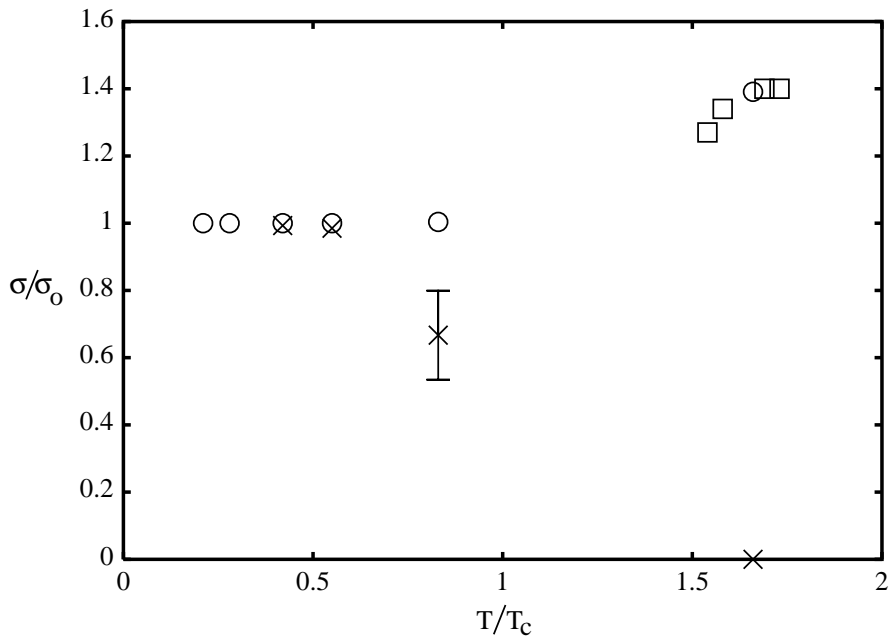}
}

\caption{String tension between two static color sources (crosses)
and spatial string tension (open symbols) as a function of temperature.
Circles denote measurements taken for $\epsilon =0, c=0.24$, whereas
squares are measurements taken at other points on the
$T_c /\sqrt{\sigma_{0} } \approx 0.69$ trajectory, namely for $N_t =1$ at
$\epsilon =0.4, 0.2, -0.2, -0.4$ (in order of ascending $T/T_c $),
cf. Table \ref{table0}.
All but one of the measurements displayed have errors smaller than the
symbols by which they are represented and result from fitting linear
functions to the extracted potentials at each temperature; this
described well all data taken at the temperature in question. The
exception is the measurement of the string tension between two color
sources at $0.83 T_c $. There, a simple linear fit fails, and logarithmic
and/or Coulomb terms were added to fit the data taken. It should be
noted that logarithmic terms are indeed known to play a role in the
static quark potential at finite temperatures and they become appreciable
near $T=T_c $, cf. {\protect \cite{kacz} }. While such a more general 
ansatz allows an acceptable fit, the string tension extracted depends on 
the precise ansatz to the extent indicated by the error bar in the figure.}

\label{fig3}

\end{figure}

Fig. \ref{fig3} displays measurements of the string tension between two
static color sources, evaluated using Polyakov loop correlators, and also
of the spatial string tension extracted from spatial Wilson loops,
at different temperatures. These measurements quantitatively reproduce
the behavior found in full Yang-Mills theory. While the correct relation
between zero-temperature string tension and deconfinement temperature
was fitted using the freedom in the choice of coupling constants, cf.
the previous section, the behavior of the spatial string tension 
constitutes a first prediction of the model. As evidenced in 
Fig. \ref{fig3}, this prediction is stable to within 10 \% effects
as one varies the point on the physical 
$T_c /\sqrt{\sigma_{0} } \approx 0.69$ trajectory at which the measurement
is taken. For the choice (\ref{choice}) of coupling constants,
the value $\sigma_{s} (T=1.67 T_c ) = 1.39 \sigma_{0} $
measured in the present vortex model agrees to within 1\% with the value
obtained in full Yang-Mills theory, as can be inferred by interpolating 
the values quoted in \cite{karsch}. A possible way to understand this
surprisingly accurate correspondence, based on the specific space-time
structure of the vortex configurations in the deconfined phase,
will be discussed further below.

In order to understand and interpret the confinement properties displayed
in Fig. \ref{fig3}, it is useful to contrast them with the percolation
properties of the vortices \cite{tlang}. To best exhibit the latter,
it is necessary to consider three-dimensional slices of the lattice
universe (where the slices are taken {\em between} two parallel hyperplanes
of the lattice on which the vortices are defined). On such slices, vortices
form closed lines made up of links. Both time slices as well as slices in
which one of the spatial coordinates is fixed (in the following referred to
as space slices) are of interest. By finding an initial vortex
link, identifying all vortex links connected to it, and iterating
this procedure with all new vortex links reached, one can discriminate
between disjoint vortex clusters; the space-time extension of a cluster
is then defined as the maximal Euclidean distance between two points on
that cluster. In a percolating ensemble, most vortex links will be
organized into clusters of near maximal size, which corresponds to an
extension of $\sqrt{N_t^2 +2 N_s^2 } /2 $ lattice spacings on a
$N_t \times N_s \times N_s $ space slice due to the periodic boundary
conditions (and analogously for time slices). On the other hand,
in the absence of percolation, most vortex links will be organized
into small clusters.

\begin{figure}[h]

\centerline{
\hspace{2cm}
\epsfxsize=9cm
\epsffile{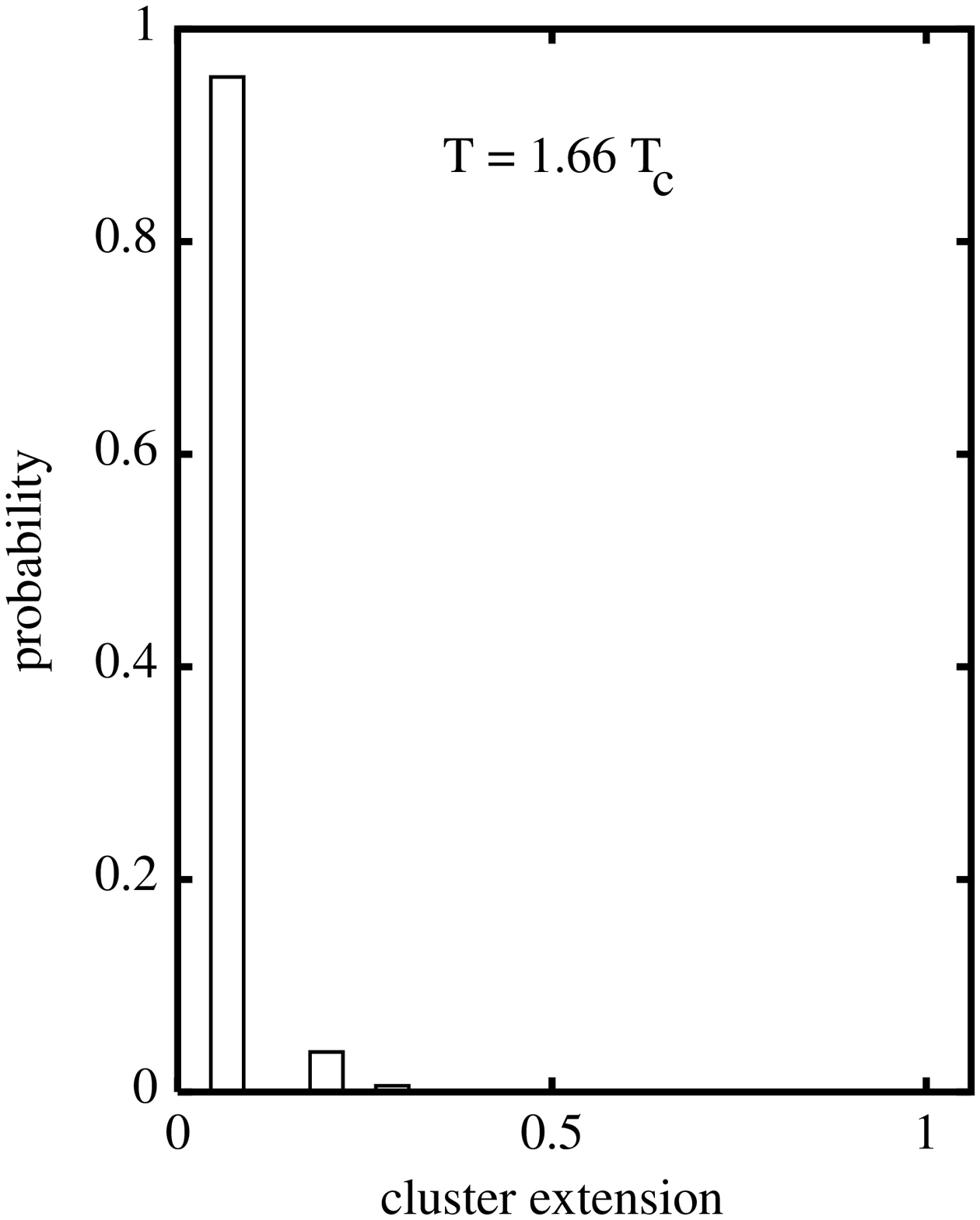}
\hspace{-1cm}
\epsfxsize=9cm
\epsffile{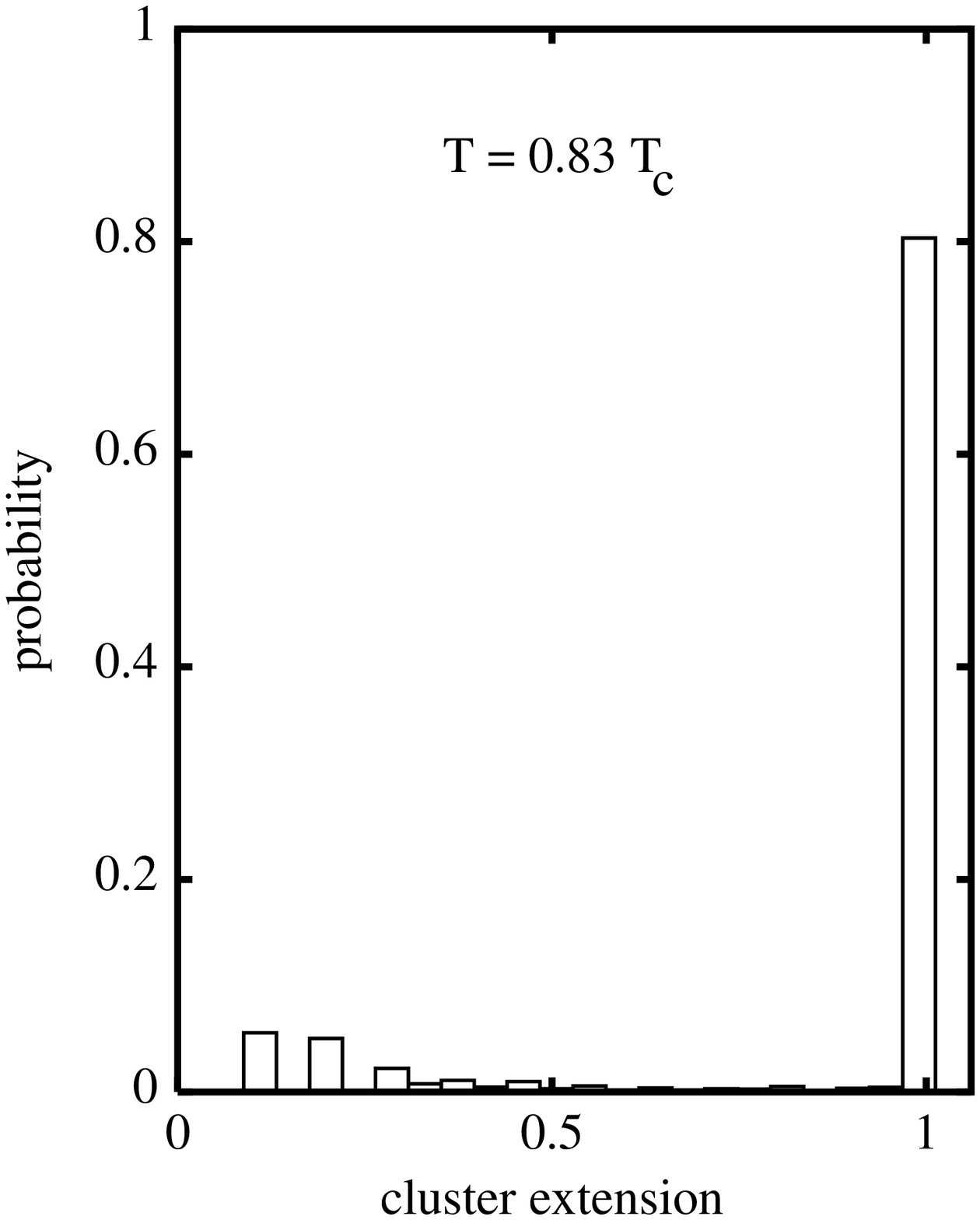}
}
\vspace{-3cm}

\centerline{
\hspace{2cm}
\epsfxsize=9cm
\epsffile{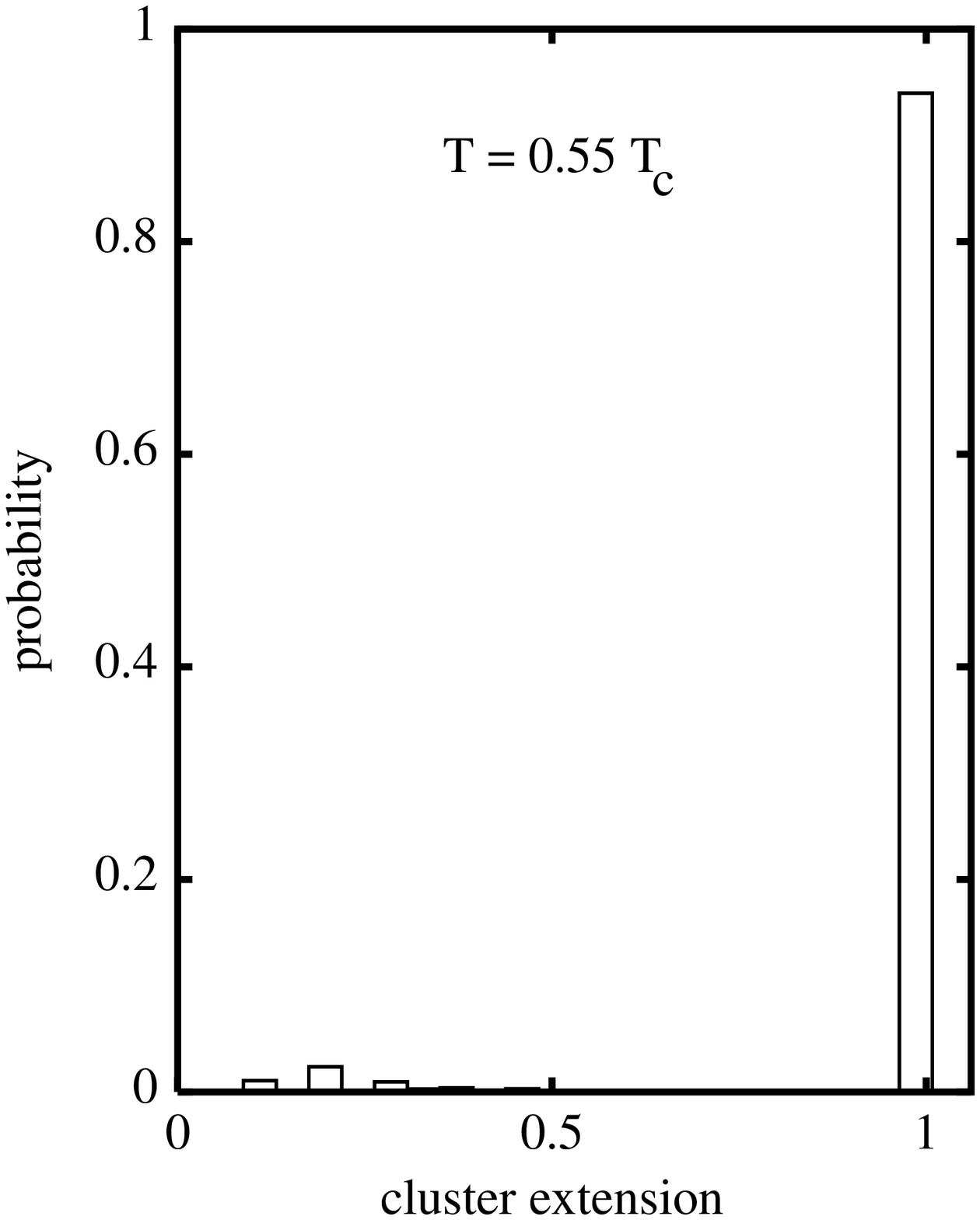}
\hspace{-1cm}
\epsfxsize=9cm
\epsffile{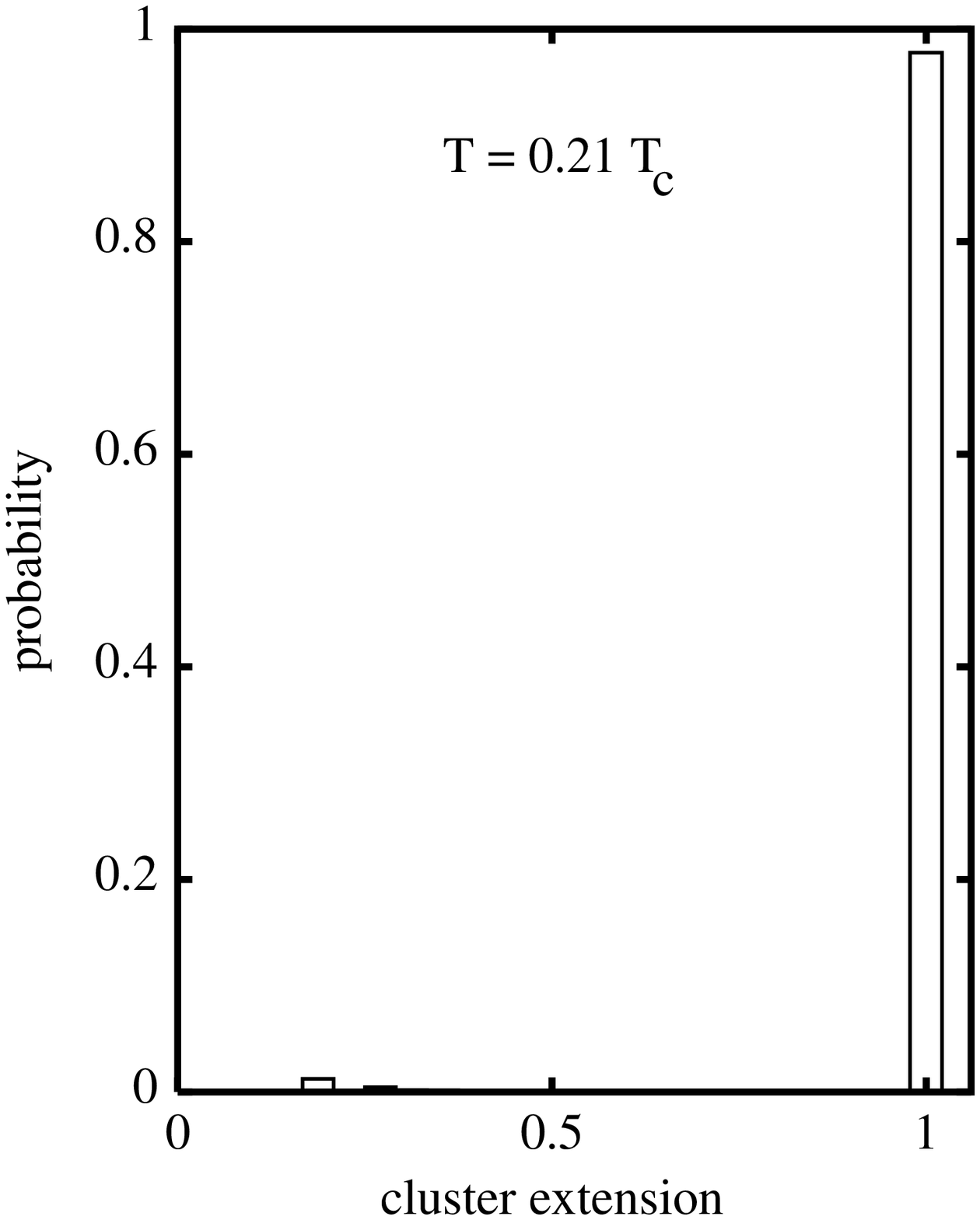}
}

\vspace{-2.7cm}

\caption{Vortex links found in space slices of the random vortex
surfaces, binned according to the extension of the cluster to which
they belong (given in units of the maximal possible extension
$\sqrt{N_t^2 +2 N_s^2 } a/2 $). The distributions were normalized to
unity, so that each bar represents the fraction of vortex links found
in clusters of the corresponding extension.}
\label{fig4}
\end{figure}

\begin{figure}[ht]
\centerline{
\epsfxsize=12cm
\epsffile{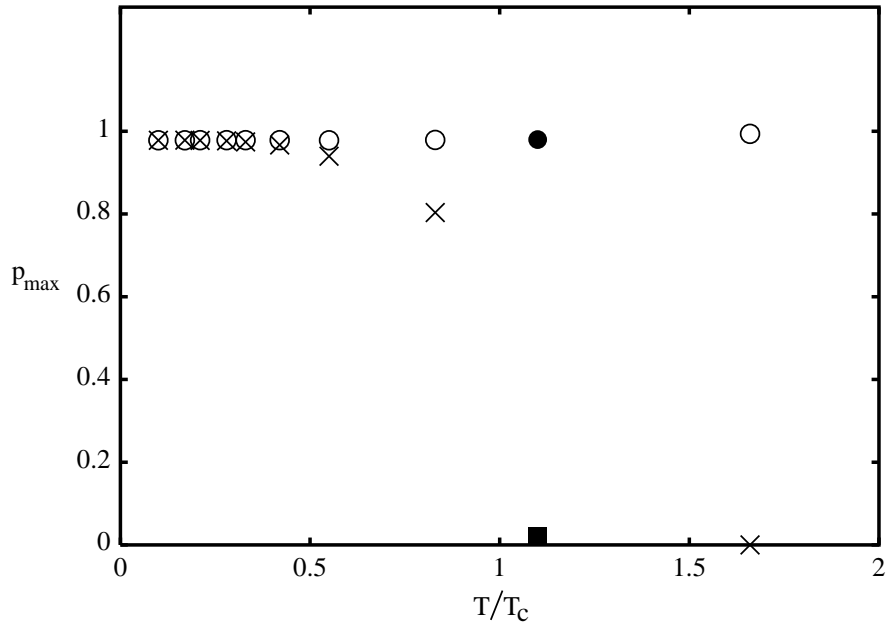}
}

\caption{Fraction $p_{max} $ of vortex links which belong specifically to
clusters whose extension lies within (roughly) 5\% of the maximal possible
extension (cf. the width of the bins in Fig. \ref{fig4}), as a function of
temperature. Open circles show $p_{max} $ measured in time slices of the
random vortex surfaces present in the ensemble, crosses show for comparison
the values found in space slices (i.e. the crosses correspond to the heights
of the bins of maximal cluster extension in Fig. \ref{fig4}). The
measurements at $T=1.1 T_c $ (full circle representing time slice and
square representing space slice) were obtained by an interpolation
procedure, see text. Evidently, clusters in time slices of the lattice
universe percolate even in the deconfined phase. Note how the behavior
of $p_{max} $ parallels the behavior of the corresponding string tensions,
cf. Fig. \ref{fig3}.}

\label{percmax}

\end{figure}

Fig. \ref{fig4} displays, for different temperatures and
taking space slices of the lattice universe, the fraction of vortex
links present in the ensemble which are part of a cluster of the
extension specified on the horizontal axis. Clearly, in the confining
phase, vortices in space slices percolate, whereas they cease to percolate
in the deconfined phase. A different picture is obtained when considering
time slices, cf. Fig. \ref{percmax}. There, vortex lines percolate in both
phases. This in particular also implies that the two-dimensional vortex
surfaces in four-dimensional space-time percolate in both phases. Only
when considering space slices of the lattice universe does a percolation
transition become visible. 

Measurements at fixed $\epsilon ,c$ (and therefore at fixed lattice spacing
$a$) only provide a discrete set of temperatures; in the case of
$\epsilon =0,c=0.24$ treated above, $N_t =2$ happens to correspond to 
$T=0.83 T_c $, whereas $N_t =1$ corresponds to $T=1.67 T_c $ (the reader
may feel uneasy at this point because $N_t =1$ seems rather special, and
the observed phenomena in the deconfined phase may be tied to this 
particular case; this concern is addressed in more detail further below). 
Clearly, it is desirable to have a better temperature resolution especially 
in the region of the phase transition. Values of observables at intermediate 
temperatures can be defined via an interpolation procedure such as already 
employed in section \ref{survsect}. This is how the values at $T=1.1 T_c $ in
Fig. \ref{percmax} were obtained. The measurements in section \ref{survsect}
in particular allowed to interpolate, for fixed $\epsilon =0$, the
value of $T_c a$ as a function of $c$. This, however, also permits
finding the set of $c_i $ for which $1.1=(Ta)/(T_c a)=1/(iT_c a)$, with
$i$ denoting an integer. Then, by construction, measurements of an
observable on a lattice with $N_t =i$ at coupling $c=c_i $ all correspond
to $T=1.1 T_c $, which again by interpolation allows to define the
observable also for $c=0.24$ at $T=1.1 T_c $.

A more specific picture of the deconfined phase is obtained by analyzing
the number of links contained in the clusters. For the choice of coupling
constants $\epsilon =0 ,c=0.24$, (only) the lattice universe
with $N_t =1$ realizes the deconfined phase. On space slices of this
lattice, clusters containing only one vortex link are necessarily clusters
which wind around the lattice in the Euclidean time direction and are
closed by virtue of the periodic boundary conditions. The smallest
non-winding vortex cluster by contrast contains four links. Indeed,
measuring the percentage of links belonging to clusters containing only
one link yields a fraction of 95\% (for $N_t =1$, i.e. $T=1.67 T_c $).
Thus, the small extension vortex clusters which dominate the deconfined
phase can more specifically be characterized as winding vortex
configurations.

Note that this specific space-time structure of the vortex configurations
in the deconfined phase, which is also found for the center projection
vortices extracted from the SU(2) Yang-Mills ensemble, cf. \cite{tlang},
may explain the surprising accuracy of the prediction of the spatial
string tension displayed in Fig. \ref{fig3}. Given that the Yang-Mills
dynamics favors the formation of winding vortices extending predominantly
into the Euclidean time direction, the vortex surfaces allowed in the
present model can quite accurately represent the configurations relevant
in the full theory, despite the large lattice spacing. Thus, in this
particular setting, the model space does not imply a strong truncation
of the full physics, even on the coarse lattice used.

The reader may feel uneasy about the discussion in the previous paragraphs
because the lattice with $N_t =1$ realizing the deconfined phase seems a
rather special case, and one might worry that the loss of percolation, and
more specifically the dominance of winding vortices, is a particular feature
of the $N_t =1$ lattice. In order to show that, on the contrary, the
correlation between deconfinement on the one hand and the dominance of
winding vortices on the other hand is a generic feature of the random vortex
surfaces, the authors have numerically investigated the (unrealistic)
choice of parameters $\epsilon =0, c=0.46$. In this case, lattices
with $N_t =1,2,3$ realize the deconfined phase, whereas lattices with
larger $N_t $ realize the confined phase (as determined by measurements
of the string tension). Table \ref{table1} displays the resulting
fraction $p_i $ of vortex links found in clusters containing a total
of $i$ links.

For $N_t =1$, the winding vortices containing one link
dominate the ensemble to more than 99.5\%. For $N_t =2$, at least
77\% of links detected were part of a winding vortex containing a
total of two links; the fraction $p_4 $ contains additional winding
vortices (with a transverse fluctuation), but the analysis is ambiguous,
since there are also non-winding clusters containing four links.
For $N_t =3$, this ambiguity does not occur, and the fractions $p_3 $,
$p_5 $ and $p_7 $ necessarily correspond to winding vortices, with
transversal fluctuations in the cases of $p_5 $ and $p_7 $. Thus, winding
vortices still encompass well more than half of the vortex links
present in the ensemble even quite near the critical temperature
(additional winding vortices are subsumed in the further odd fractions
$p_9 , p_{11} ,\ldots $ not shown). Percolating clusters (embodied in
$p_{max} $) on the other hand are virtually non-existent for the
deconfined cases $N_t =1,2,3$. By contrast, for $N_t \geq 4$,
which corresponds to the confining phase, the dominant proportion of
vortex links is associated with percolating vortices (cf. $p_{max} $).
Short winding vortices completely disappear, including ones with
small transverse fluctuations (cf. $p_5 $ and $p_7 $ in the case
$N_t =5$).

\begin{figure}[h]

\centerline{
\hspace{1.8cm}
\epsfxsize=9cm
\epsffile{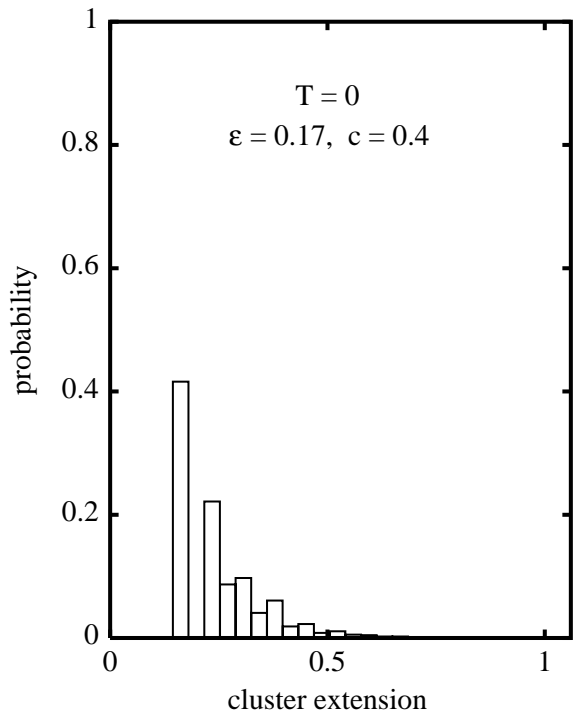}
}
\vspace{-2.7cm}

\caption{Vortex links found in a three-dimensional slice of a $16^4 $
lattice universe, binned according to the extension of the cluster to 
which they belong (given in units of the maximal possible extension
$\sqrt{3\cdot 16^2 } a/2 $). Measurements were taken in the ensemble
generated by the choice of coupling constants $\epsilon =0.17, c=0.4$
belonging to the non-confining regime, cf. Fig. \ref{fig2}.
The distributions are normalized to unity, so that each bar represents 
the fraction of vortex links found in clusters of the corresponding 
extension.}

\label{fig7}
\end{figure}

As a last point, it is interesting to contrast the percolation phenomena
exhibited above with the behavior of the vortices in the region of the
phase diagram in which confinement is absent even at zero temperature
(the shaded region in Fig. \ref{fig2}). In the cases studied further
above, taken from the confining regime of the coupling constant plane, a
percolation transition at the deconfinement temperature $T_c $ only became
visible in space slices of the lattice universe; on the other hand,
percolation of vortex lines in time slices, and consequently percolation of
the complete vortex surfaces in four dimensions, persisted even above the
deconfinement temperature $T_c $. In contradistinction to this, in the
non-confining regime of the coupling constant plane, the two-dimensional
vortex surfaces in four-dimensional space-time as a whole do not percolate.
Note that this is equivalent to the statement that vortex lines do not
percolate in {\em any} three-dimensional slice of space-time (at zero
temperature, or any symmetric lattice approximation of it, there is
of course no distinction between space and time slices). As a case in
point, therefore, Fig. \ref{fig7} displays the sliced vortex distribution
analogous to Fig. \ref{fig4}, but evaluated using a symmetric ($16^4 $)
lattice for $\epsilon =0.17, c=0.4$, which is in the non-confining region,
cf. Fig. \ref{fig2}.
\vspace{0.7cm}

\begin{table}

\[
\begin{array}{|c||c|c|c|c|c|c|c|c|}
\hline
N_t & p_1 & p_2 & p_3 & p_4 & p_5 & p_6 & p_7 & p_{max} \\
\hline\hline
1 & 1.00 & 0 & 0 & 0 & 0 & 0 & 0 & 0 \\ \hline
2 & 0 & 0.77 & 0 & 0.14 & 0 & 0.06 & 0 & 0 \\ \hline
3 & 0 & 0 & 0.41 & 0.07 & 0.08 & 0.06 & 0.05 & 0 \\ \hline\hline
4 & 0 & 0 & 0 & 0.04 & 0 & 0.02 & 0 & 0.84 \\ \hline
5 & 0 & 0 & 0 & 0.02 & 0 & 0.01 & 0 & 0.89 \\ \hline
6 & 0 & 0 & 0 & 0.02 & 0 & 0.01 & 0 & 0.91 \\ \hline
7 & 0 & 0 & 0 & 0.02 & 0 & 0.01 & 0 & 0.92 \\ \hline
8 & 0 & 0 & 0 & 0.02 & 0 & 0.01 & 0 & 0.92 \\ \hline
\end{array}
\]
\vspace{0.3cm}

\caption{Fraction $p_i $ of vortex links found in clusters containing
a total of $i$ links, as measured by taking space slices of the 
$\epsilon =0, c=0.46$ random vortex surface ensemble on lattices
with a temporal extension $N_t a$. For these values of the coupling
constants, $N_t =1,2,3$ realize the deconfined phase, whereas higher
$N_t $ correspond to the confining phase. For comparison, the table
also quotes the fraction $p_{max} $ of links found in clusters whose 
extension lies within (roughly) 5\% of the maximal possible space-time 
extension (cf. the width of the bins in Fig. \ref{fig4}).}

\label{table1}

\end{table}

\section{Discussion}
\begin{figure}[t]
\centerline{
\epsfxsize=4cm
\epsffile{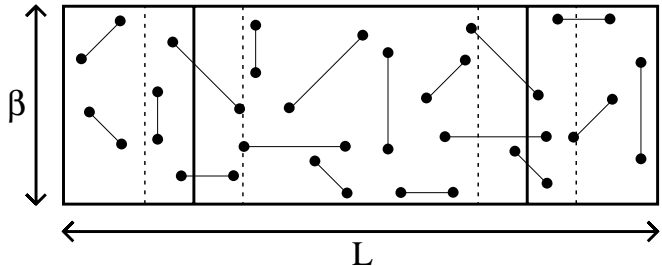}
}
\vspace{-1cm}
\caption{Pairs of vortex intersection points of maximal separation $d$
on a space-time plane. The thin lines connecting the points are merely
to guide the eye in identifying the pairs. The existence of a maximal 
separation is a consequence of the vortices generating the intersection 
points being organized into clusters of maximal extension $d$ in a higher
dimension, i.e. these clusters do not percolate. Only pairs whose
midpoints lie within the strips of width $d$ (delineated by the 
interrupted lines) centered on the two Polyakov loops (solid lines)
making up a Polyakov loop correlator have the possibility of 
contributing a factor $(-1)$ to the latter.}
\label{heur}
\end{figure}

It is not surprising that the high temperature, deconfined phase of the
vortex model is associated with a lack of vortex percolation in space 
slices of the universe. For any Polyakov loop correlator, one can choose 
a space slice containing both of the Polyakov loops involved as well as 
the minimal area spanned by them. Consider now the consequence of a lack 
of vortex percolation in this space slice in the following simple heuristic 
picture.  Absence of percolation implies the existence of an upper bound 
$d$ on the size of vortex clusters. Due to the closed nature of the vortex
lines, this implies that, on the space-time plane containing the two 
Polyakov loops, any point at which a vortex pierces that plane comes
paired with another such point at most a distance $d$ away, 
cf. Fig. \ref{heur}. 

Consider the idealized case of such pairs
being randomly distributed on the space-time plane in question. Then
one can evaluate the behavior of the Polyakov loop correlator at inverse
temperature $\beta $ on a universe of linear extension $L$ as
follows. Only pairs whose midpoints lie within the two strips of
width $d$ centered on the Polyakov loops can contribute a factor $-1$ to
the Polyakov loop correlator. Denote by $p$ the probability that such
a pair actually does contribute a factor $-1$. This probability is an 
appropriate average over the distances of the midpoints of the pairs from 
the Polyakov loops, their angular orientations and the distribution of 
separations between the points making up the pairs. The probability $p$, 
however, does not depend on the macroscopic extension of the Polyakov
loop correlator. Now, a pair which is placed at random on the space-time
plane has probability $p\cdot A/\beta L $ of contributing a factor $-1$ to
the Polyakov loop correlator, where $A=2 \beta d$ is the area of the
two strips of width $d$ centered on the Polyakov loops, and $\beta L$
is the area of the entire plane. Placing $N_{pair} $ pairs on the plane
at random, the probability that $n$ of them contribute a factor $-1$ to 
the Polyakov loop correlator is
\begin{equation}
P_{N_{pair} } (n) = \left( \begin{array}{c} N_{pair} \\ n \end{array} \right)
\left( \frac{2pd}{L} \right)^{n}
\left( 1 - \frac{2pd}{L} \right)^{N_{pair} -n}
\end{equation}
and, consequently, the expectation value of the correlator for large
universes is
\begin{equation}
\langle W \rangle = \sum_{n=0}^{N_{pair} } (-1)^{n} P_{N_{pair} } (n)
= \left( 1-\frac{4pd}{L} \right)^{N_{pair} }
\stackrel{N_{pair} \rightarrow \infty }{\longrightarrow }
e^{-2\beta pd\rho }
\end{equation}
where the planar density of points $\rho =2N_{pair} /\beta L $ is kept
fixed as $N_{pair} \rightarrow \infty $. The Polyakov loop correlator is 
therefore independent of the separation between the Polyakov loops, 
negating confinement, if the extension of vortices or vortex networks 
in a space slice of the universe is bounded. Note that the persistence 
of percolation of the two-dimensional vortex surfaces as a whole in the 
deconfined phase does not influence this argument; what is important is 
the presence or absence of a pair correlation between vortex intersection 
points on the plane containing a Polyakov loop correlator.

Conversely, percolation of vortices is therefore a necessary condition
for confinement. Only then is it possible for the points at which
vortices pierce a given space-time plane to be sufficiently randomly 
distributed as to generate an area law for a Wilson loop or Polyakov
loop correlator embedded in that plane; the pair correlation crucial
in the model visualization presented above is no longer operative.
Indeed, if one assumes these piercing points to be randomly distributed,
one obtains by an argument analogous to the one above an area law with 
a string tension equal to twice the planar density of intersection 
points $\rho $, cf. \cite{corr}. More generally, if the points cannot 
be packed arbitrarily densely, but instead at most one per plaquette of 
a lattice (of spacing $a$) imposed on the plane can occur, the string 
tension obeys \cite{giedt}
\begin{equation}
\sigma /\rho = -\frac{1}{\rho a^2 } \ln (1-2\rho a^2 )
\label{rplat}
\end{equation}
which reduces to the value $\sigma /\rho =2$ quoted above in the
limit $a\rightarrow 0$.

The relation between the planar vortex density $\rho $ and the string
tension arising in the simple random picture (\ref{rplat}) is obeyed
to a good approximation by the values measured at zero temperature
in the present vortex model (with $\epsilon =0, c=0.24$). At zero 
temperature, one has $\sigma a^2 = 0.755$ and $\rho a^2 = 0.27 $,
which fulfills (\ref{rplat}) up to a 3\% deviation; the measured
quantities are thus consistent with a random distribution of vortex
intersection points on any given space-time plane. The same behavior
is found for the center projection vortices of Yang-Mills theory
{\em after} subjecting them to a smoothing procedure \cite{bertle};
the density of unsmoothed center projection vortices is significantly
higher \cite{temp}. This is consistent with the physical interpretation
discussed in section \ref{sectphys}. Quantitatively, the model vortex
density quoted above differs from the center projection vortex density
in SU(2) Yang-Mills theory \cite{temp} by a factor two.

The necessity of percolation for an area law behavior of the Wilson loop
furthermore explains the persistence of percolation in time slices in 
the deconfined phase. Otherwise, spatial Wilson loops, which can be
embedded in a time slice, could not continue to obey an area law above
the deconfinement temperature. On the other hand, spatial Wilson loops
can also be embedded in space slices, in which percolation ceases in
the deconfined phase. The reason one can nevertheless still understand 
the spatial string tension in the space slice picture lies in the
different topological setup: Vortices winding in the Euclidean time
direction, which dominate the deconfined phase, pierce spatial Wilson
loops at isolated points despite being of limited extension. The
pair correlation between the piercing points which would preclude an
area law does not arise due to the possibility of closing a short
vortex line via the periodic boundary conditions.

To sum up, in the vortex picture there is a strong connection
between confinement and the percolation properties of vortices.
Percolation is a necessary condition for confinement; the deconfinement
transition is induced by a percolation transition to a phase which
lacks percolating vortex clusters (when an appropriate slice of
the configurations is considered). Also in this respect, the behavior
of the vortex model presented here closely parallels the behavior
found for the center projection vortices of Yang-Mills theory \cite{tlang}.
This connection between percolation and confinement moreover is
one of the points at which the duality between the (magnetic) vortex
picture and electric flux models becomes apparent. In electric flux
models, the deconfinement transition also takes the guise of a percolation
transition \cite{patel}; however, it is the {\em deconfined} phase in which 
electric flux percolates.

While this clarifies {\em how} vortex configurations generate the
confined and deconfined phases, the simple structure of the vortex model 
investigated in this work also allows an intuitive understanding of the 
underlying dynamics, i.e. {\em why} the vortices behave as they do.
Qualitatively, the parameters entering the vortex action have the following
effects. The action per plaquette area $\epsilon $ (cf. eq. (\ref{epsdef}))
acts as a chemical potential for the mean density of vortices, whereas the 
curvature penalty $c$ (cf. eq. (\ref{cdef})) imposes an ultraviolet cutoff 
on the space-time fluctuations of the vortex surfaces. To a certain extent, 
the two effects can be traded off with one another; striking evidence of 
this is provided by the approximately invariant physics found on the 
$T_c /\sqrt{\sigma_{0} } \approx 0.69$ trajectory depicted in 
Fig. \ref{fig2}. This can be understood as follows: If one generates two 
vortex configurations at random, then in all but exceptional cases, the 
configuration with the higher mean vortex density will also contain the 
higher amount of total curvature. Therefore, both coupling constants 
$\epsilon $ and $c$ simultaneously curtail both the mean vortex density 
and the curvature. If one raises either $\epsilon $ or $c$, the mean 
density of vortices falls (and, along with it, the zero-temperature string 
tension). This gradual decrease of the string tension with the density does 
not persist indefinitely; instead, at some point, the vortex density becomes 
so low that the vortex structures lose their connectivity and form isolated 
clusters instead of percolating throughout space-time, cf. Fig. \ref{fig7}. 
This implies a pair correlation between vortex intersection points on a 
space-time plane which leads to an immediate loss of confinement, as 
already discussed further above. Therefore, despite there remaining a 
finite vortex density, the string tension vanishes; in this way, the 
confining and non-confining regions in Fig. \ref{fig2} are generated.

Turning to the case of finite temperatures, the deconfining dynamics
in the random vortex model can be understood in terms of an entropy 
competition. As one shortens the (lattice) universe in the Euclidean 
time direction, a new class of vortex clusters of small extension
(viewed in a space slice) becomes available, namely the winding
vortices which have been verified above to dominate the deconfined
phase. Thus, the entropy balance between the class of percolating
vortex configurations and the class of limited extension, non-percolating
configurations shifts towards the latter. This interpretation of the
deconfining dynamics is almost a tautology in view of the simple
formal structure of the vortex model presented here. Evaluating
the partition function (cf. eq. (\ref{partit})) of the model by 
construction amounts precisely to enumerating all possible closed
surface configurations, given a certain mean vortex density (enforced
by the coupling constants $\epsilon $ and, indirectly, $c$) and an
ultraviolet cutoff on the fluctuations of the vortex surfaces
(embodied in the lattice spacing $a$ and reinforced by the curvature
penalty $c$). No other dynamical information enters the model, and
therefore it can be nothing but the entropy associated with the
different classes of random surfaces which determines which phase is
realized.

\section{Outlook}
In the previous sections, a model of infrared Yang-Mills dynamics
was presented which allows an intuitive understanding of both the
confinement phenomenon and the transition to a deconfined phase
at finite temperatures. These properties are closely tied to the
percolation characteristics of the vortex surfaces on which the
model is based. The behavior of the random surface ensembles
generating the two phases closely parallels the behavior found for 
center projection vortices in the Yang-Mills ensemble \cite{tlang}. 
It is possible to choose the coupling constants of the vortex model such
that long-range static quark potentials and spatial string tensions 
measured in Yang-Mills theory are quantitatively reproduced at all 
temperatures up to the cutoff of the model. The correct description 
of the spatial string tension in the deconfined phase should be noted 
in particular; this nontrivial feature at no point entered either the 
construction of the model or the choice of coupling constants.

The vortex model presented here was formulated to describe Yang-Mills
theory with an SU(2) color group. The case of SU(3) color realized 
in nature will in some respects exhibit qualitatively different behavior; 
the center vortex model appropriate for this gauge group is currently
under investigation. Since there are two nontrivial center elements in the 
SU(3) group, namely the phases $e^{\pm i 2\pi /3} $ (multiplied by the 
$3\times 3 $ unit matrix), one must allow for two distinct vortex fluxes.
The main qualitative difference in the topology of vortex configurations
is the presence of vortex branchings; a vortex carrying one type of vortex 
flux can split into two vortices carrying the other type of vortex flux,
as long as flux conservation is respected.

On the other hand, to provide a comprehensive picture of the infrared 
sector of QCD, the vortex model must also be investigated with a view to
describing the topological susceptibility of the Yang-Mills ensemble
and the spontaneous breaking of chiral symmetry. The manner in which
vortex configurations generate a nontrivial Pontryagin index was
recently clarified in \cite{cont}; the relevant properties are encoded
in the (oriented) self-intersection number of the vortex surfaces.
In a companion paper \cite{preptop}, those results are implemented on
a space-time lattice, allowing a measurement of the topological 
susceptibility. Also for this quantity, the vortex model generates
a realistic value, compatible with lattice measurements in full SU(2) 
Yang-Mills theory. In the same vein, it is necessary to construct efficient 
ways to evaluate the spectrum of the Dirac operator in a vortex 
background. This will make it possible to calculate the associated 
chiral condensate \cite{cashb}, which represents an order parameter 
for the spontaneous breaking of chiral symmetry. 

As a last remark, it is tempting to speculate that the phase diagram
of electroweak theory \cite{phipsen} can similarly be understood in 
terms of the percolation characteristics of electroweak vortices
\cite{ilge}, in particular as far as the confinement properties are 
concerned. In the Higgs phase, the coupling to the Higgs condensate may 
penalize the vortex density to such an extent that the theory enters the 
non-percolating, non-confining regime (the shaded region in 
Fig. \ref{fig2}), thus allowing electroweak gauge bosons to be seen as 
asymptotic states.

\section{Acknowledgements}
M.E. thanks B.Petersson and G.Thorleifsson for an illuminating discussion
on random surface models.

\end{document}